\documentclass[12pt,preprint]{aastex}

\shorttitle{Stuctural Parameters of GCs in NGC 5128}
\shortauthors{G\'{o}mez and Woodley}

\begin{document}

\title{Sizes of Confirmed Globular Clusters in NGC 5128: A Wide-Field High-Resolution Study\footnote{This paper includes data gathered with the 6.5 meter Magellan Telescopes located at Las Campanas Observatory, Chile}}

\author{Mat\'{\i}as G\'{o}mez}
\affil{Grupo de Astronom\'{\i}a, Depto. de F\'{\i}sica, Universidad de
  Concepci\'{o}n, Casilla 160-C, Concepci\'{o}n, Chile}
\email{matias@astro-udec.cl}

\author{Kristin A.~Woodley}
\affil{Department of Physics \& Astronomy, McMaster University,
  Hamilton ON L8S 4M1, Canada}
\email{woodleka@physics.mcmaster.ca}

\begin{abstract}
Using Magellan/IMACS images covering a 1.2 x 1.2 sq. degree
FOV with seeing of 0.4"-0.6", we have applied convolution techniques 
to analyse the light distribution of 364 confirmed globular cluster 
in the field of NGC 5128 and to obtain their structural parameters. Combining
these parameters with existing Washington photometry from 
Harris et al. (2004), 
we are able to examine the size difference between metal-poor (blue) 
and metal-rich (red) globular clusters.
For the first time, this can be addressed on a sample of confirmed
clusters that extends to galactocentric distances about 8 times 
the effective radius, R$_{eff}$, of the galaxy.
Within 1 R$_{eff}$, red clusters are about $30\%$ 
smaller on average than blue clusters, in agreement
with the vast majority of extragalactic globular cluster systems
studied.  As the galactocentric distance increases,
however, this difference becomes negligible. Thus, our results 
indicate that the difference in the clusters' effective radii, r$_e$, could be
explained purely by projection 
effects, with red clusters being more centrally concentrated than blue
ones and an intrinsic r$_e$--R$_{gc}$ dependence, like the one 
observed for the Galaxy. 
\end{abstract}


\section{Introduction}
\label{sec:intro}
Since sizes and structural parameters of globular clusters (GCs)
in different GC systems (GCSs) have first been obtained,
it has become clear that some of these properties correlate with 
global properties of
their host galaxies \citep[see for example][]{jordan05,brodie06}.
The existence of the so called fundamental plane relation for
an increasing number of studied GCSs seems to
confirm that GCs populate
a narrow region in this parameter space
\citep{djorgovski95,mclaughlin00,mclaughlin05,barmby07}.
However, there are puzzling trends that are still awaiting confirmation
and need to be addressed using larger samples of GCs. 

It is necessary to study structural parameters of 
GCs and GC-like objects in different
environments before definitive statements can be made 
regarding their formation.
Among the structural parameters that can be studied, the effective
(or half-light) radius is of particular importance.
Models have shown that this
quantity remains fairly constant throughout the entire GC 
lifetime \citep{spitzer72,aarseth98}, making it a
good indicator of proto-GC sizes that are still observable today. 
A decade ago, HST observations unveiled a systematic size difference 
between red and blue GCs \citep{kundu98}.  Since then, multiple 
studies have found that the
blue GCs are between $17\%-30\%$ larger than their
metal-rich counterparts in both spirals and early--type galaxies
\citep{kundu99,puzia99,larsen01,larsen_fb01,kundu01,barmby02,jordan05}.  
However, most of these studies have made use of HST observations 
and examine only the innermost regions of the galaxy or small
fields in regions at galactocentric distances 
greater than the galaxy's effective radius.

According to \cite{larsen03}, the systematic size difference between
red and blue GCs is caused merely by a projection effect.
Since red (metal-rich) GCs are found to be more 
centrally concentrated than blue
(metal-poor ones) in early type galaxies 
\citep[][among others]{cote01,dirsch03,woodley05}, the red 
GCs will appear to lie, on average,
at a smaller galactocentric distance.  The red clusters will on
average be smaller than the blue clusters assuming that both types 
shares the same relation between the GC size and galactocentric
distance.  The relation r $\sim \sqrt{\rm{R}_{gc}}$ was first found in the
Milky Way by \cite{vandenbergh91}.  In this scenario, the difference
between the cluster sizes should be most apparent at small
galactocentric distance and should decrease strongly beyond 1 galaxy
effective radius \citep{larsen03}. 

Alternatively, \cite{jordan04} suggests that this effect could be
explained by an intrinsic difference between metal-rich and metal-poor
GCs. Assuming
half-mass radii that are independent of metallicity, effects of 
mass segregation combined with
a metallicity-dependent stellar lifetime should lead to different
sizes between the blue and red clusters.  The brightest stars would be
more massive and more centrally concentrated for the metal-rich
GCs.  This scenario should have little to no dependence
on a cluster's distance from the center of its parent galaxy.

In a recent study, \cite{spitler06} analysed the GCS 
of NGC 4594 (Sombrero, at a distance of $ \sim 9$ Mpc) using a 
six-image mosaic from HST/ACS. 
They confirm that within the inner 2 arcmin (2.2 R$_{eff}$), 
the metal-rich GCs are, on average, $17\%$
smaller than the metal-poor clusters. However, the
size difference becomes negligible at $\sim 3$ arcmin, corresponding
to $\sim 3.4$ R$_{eff}$, where R$_{eff} =0.89$ 
arcmin \citep{baggett98}. 

To further understand the sizes of red and blue clusters, we need a
homogeneous survey of a GCS with the ability to eliminate
contaminating sources, high resolution to measure structural
parameters, and over a large range in galactocentric distance.

NGC 5128 is the nearest giant elliptical galaxy, at a distance of 3.8 Mpc
\citep{mclaughlin07}.  Its GCs are thus
easily resolvable with sub-arcsecond seeing \citep{harris06}. 
In this paper we present effective radius
results for 337 GCs from the \cite{woodley07} catalog that are
confirmed GCs by either radial velocity measurement from various
studies \citep[see the references in ][]{woodley07} or are resolved by
HST/ACS images \citep{harris06}.  We also present the effective radii
of 27 GCs newly confirmed through radial velocity
measurements using the Baade 6.5-m telescope with the instrument LDSS-2
(data in preparation for publication). This list represents a clean 
sample of confirmed clusters.  All of these also have ellipticities less than
0.4 and effective radii less than 8 pc, both of which are consistent with 
{\it normal} GC properties in NGC 5128.  We find that only an additional $2.4\%$
of GCs from the \cite{woodley07} catalog have effective radii greater
than the 8 pc boundary we have imposed here (to be discussed in detail 
in  G\'omez \& Woodley, 2008, in preparation).
Those few GCs are not considered here as our purpose is to
establish the effective radius trends within the bulk of the GC population.  

\section{Observations}
\label{sec:obs}

On the night of April 9, 2006, 25 fields were imaged with the
Magellan 6.5 m telescope using the Inamori Magellan Areal Camera 
and Spectrograph (IMACS).
In the highest imaging resolution, IMACS offers a FOV of 15.4 
arcmin on a side, composed of a mosaic of 8 2Kx4KCDs
with a scale of 0.111 arcsec/pixel.
Our observational material will be fully discussed in 
Harris et al. 2008 (in preparation).  The 
total field of view of our images is roughly 1.2 x 1.2
square degrees and the average seeing is about 0.5" across the 
entire field with individual
frames ranging from 0.35" to 0.7".
Images were acquired through B (on 16 of the 25 fields) and R 
(on all 25 fields) filters with both 10 second and
300 second exposures to avoid saturation of the brightest clusters.

We have identified all GCs in the catalog of
\cite{woodley07} on our IMACS frames\footnote{Note that the positions of two
GCs have been corrected: GC0001 with a corrected right ascension
of 13$^h$ 25$^m$ 1.16$^s$ (J2000) and GC0002 with corrected
declination of -43$^{o}$ 02$^{\arcmin}$ 42.9$^{\arcsec}$ (J2000).}.   
We have run the code ISHAPE \citep{larsen99,larsen_s01} individually on each 
GC in our R filtered IMACS frames, using a
stellar point spread function (PSF) modelled from the chip in which
the cluster is located.  For this, typically 20-30 stars were
chosen in each frame and measured with standard tools in IRAF\footnote{IRAF is
  distributed by the National Optical Astronomy Observatory, which is 
operated by the Association of Universities for Research in Astronomy
Inc., under cooperative agreement with the National Science Foundation.}.
ISHAPE convolves the PSF with analytical profiles and 
compares the result with the input image until a best match is achieved.

As the analytical model, we chose King (1962) profiles, given their simplicity
and because they are known to provide a good fit to a large family of
GCs in different environments. Moffat functions were also
tried, but they do not improve the fits except for a handful of large and
very elliptical sources that we are not considering for the present study. 
They will be discussed as special cases in a forthcoming paper.  
Possible systematic effects in the sizes, arising from the choice of a
particular model are discussed in \cite{larsen99}. However, the effective
radius seems to be independent of the model for sources that have a similar
extension to the stellar PSF, as in our case.
For a recent comparison between different models, the reader is referred
to \cite{barmby07} and \cite{mclaughlin07}.

\cite{king62} profiles are defined by a core-radius r$_c$ and a
concentration index, which we define here as $c =$ r$_t/$r$_c$, where 
r$_t$ is the tidal radius of
the cluster. Usually, the concentration parameter is the most uncertain
one to constrain \citep{larsen_s01}, but given the high 
spatial resolution of our IMACS images, we were able to 
fit this along with the ellipticity, position angle (PA) and r$_c$.

The sizes quoted by ISHAPE were transformed into effective radii using
the approximation r$_e/$r$_c \approx 0.547 c^{0.486}$, good to
$\pm 2\%$ for $c>4$ \citep{larsen_s01}.  The median value for the 
concentration parameter for GCs in NGC 5128 was $c=39.4 \pm 10.2.$
Uncertainties in the effective radius were estimated by the
standard deviation of the determined value using King profiles with
fixed concentration indices of 15, 30, and 100. 
These concentration parameters were chosen
based on typical values observed in our Galaxy as well as the 
concentration parameters fit freely with ISHAPE for the NGC 5128 data. 
The r$_e$ determined for any given cluster with varying concentration
parameters does not vary more than $\sim 10\%$ for the average
GC.  The concentration parameter, c, is the most uncertain
of the fitted parameters.  The extension of the GC is a secondary uncertainty. 
A GC at the distance of NGC 5128, with an effective radius of 6 pc, 
would span a diameter of 0.6", marginally larger than the typical stellar FWHM.  For 
smaller or more compact objects, the instrisic size can be as small as
0.1", i.e. completely blurred even with sub-arcsecond imaging.

\section{Results}
\label{sec:results}

\begin{figure}
\plotone{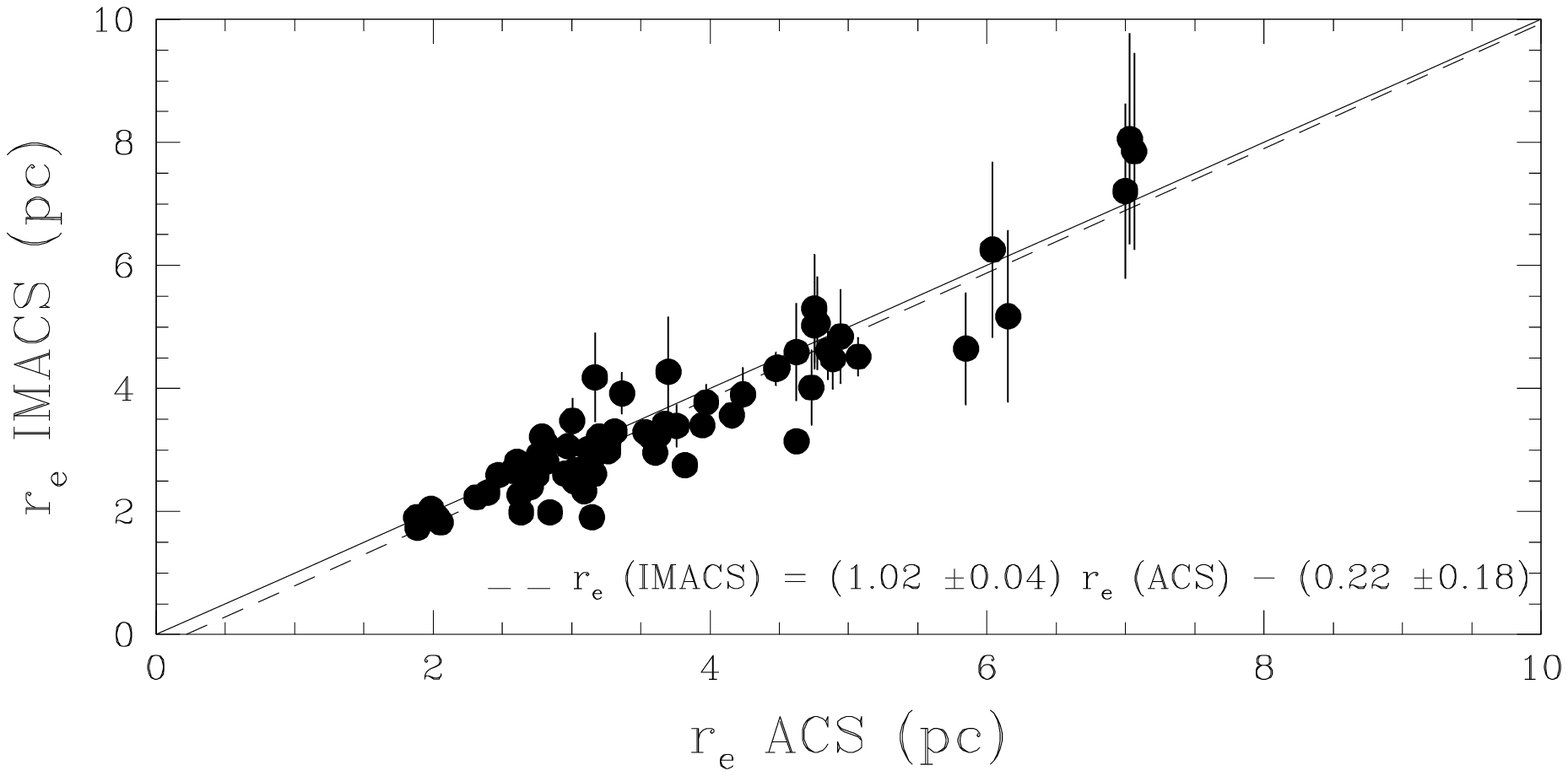}
\caption{Comparison of effective radius r$_e$, in parsecs, for 69 GCs in NGC
5128 measured with both our IMACS study and HST/ACS
\citep{mclaughlin07}.  The solid line is a 1:1 relationship.
A least squares fit (dashed line) gives 
r$_e$~(IMACS)~$ = (1.02 \pm 0.04)$~r$_e$~(ACS)~$ - (0.22 \pm 0.18)$. 
The IMACS uncertainties
correspond to the standard deviation of the effective radii in 
the ISHAPE fitting, using three different concentration 
parameters $c=15,30,100$. The agreement between these independent 
techniques and datasets is evident.} 
\label{fig:acs_imacs}
\end{figure}

We have 69 GCs in common with the HST/ACS structural parameter
study of \cite{harris06}. Their r$_e$ values, derived
through an isophotal analysis of the resolved clusters, and discussed
fully in \cite{mclaughlin07},
serve as an external comparison and quality test for our
measurements.  Figure~\ref{fig:acs_imacs} shows clearly that 
there is good agreement in the $r_e$ values determined by these independent techniques.
We have also examined our measured r$_e$ as a function
of ellipticity and luminosity and also found no notable correlation.

In an upcoming paper, (G\'omez \& Woodley, 2008, in preparation), we
will discuss the structural parameters in detail as well as the new GCs
discovered with the Baade 6.5m telescope (mentioned above) that have
been used in this study.  Here, we focus on the dependence of the 
GC sizes as a function of galactocentric radius, R$_{gc}$ for the GC 
subpopulations.  A metallicity break has been
chosen to represent the red or metal-rich ([Fe/H]$ > -1$) and blue or
metal-poor ([Fe/H]$ < -1$) subpopulations of clusters, following the
studies of \cite{larsen03,harris04,woodley05,gomez06,woodley07}.  The [Fe/H]
values were obtained from a $C-T_1$ transformation \citep{harrisharris02}
assuming E(B-V) = 0.11 \citep{schlegel98}.

\begin{figure}
\plotone{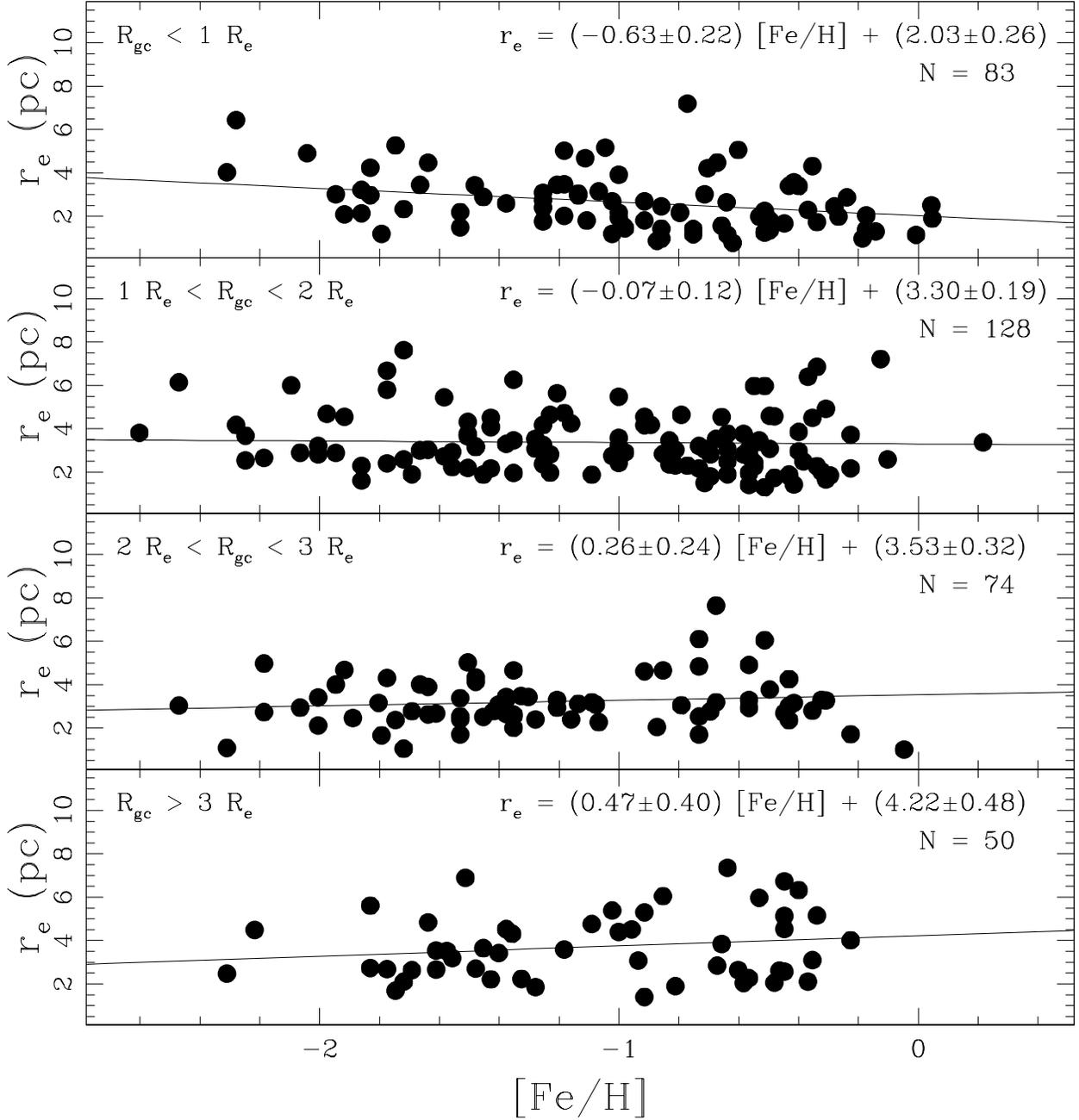}
\caption{The effective radius, r$_e$, as a function of [Fe/H] for (from top to
  bottom) R$_{gc} < 1$ R$_{eff}$ of NGC 5128, 
1 R$_{eff} <$ R$_{gc} <$ 2 R$_{eff}$, 2$ R_{eff} <$ R$_{gc} < 3 $R$_{eff}$, and
  R$_{gc} > 3$ R$_{eff}$, where 1 R$_{eff} = 5.1'$.  Least squares best fit
  and the number of GCs are indicated in each of the four
  panels.  Only GCs with ellipticity less than 0.4 and
  r$_e < 8 $ pc were considered in the fit. The change in the slope
of the fitted lines is due primarily to an increase in the median size of red
clusters.}
\label{fig:re_feh}
\end{figure}

\begin{figure}
\plotone{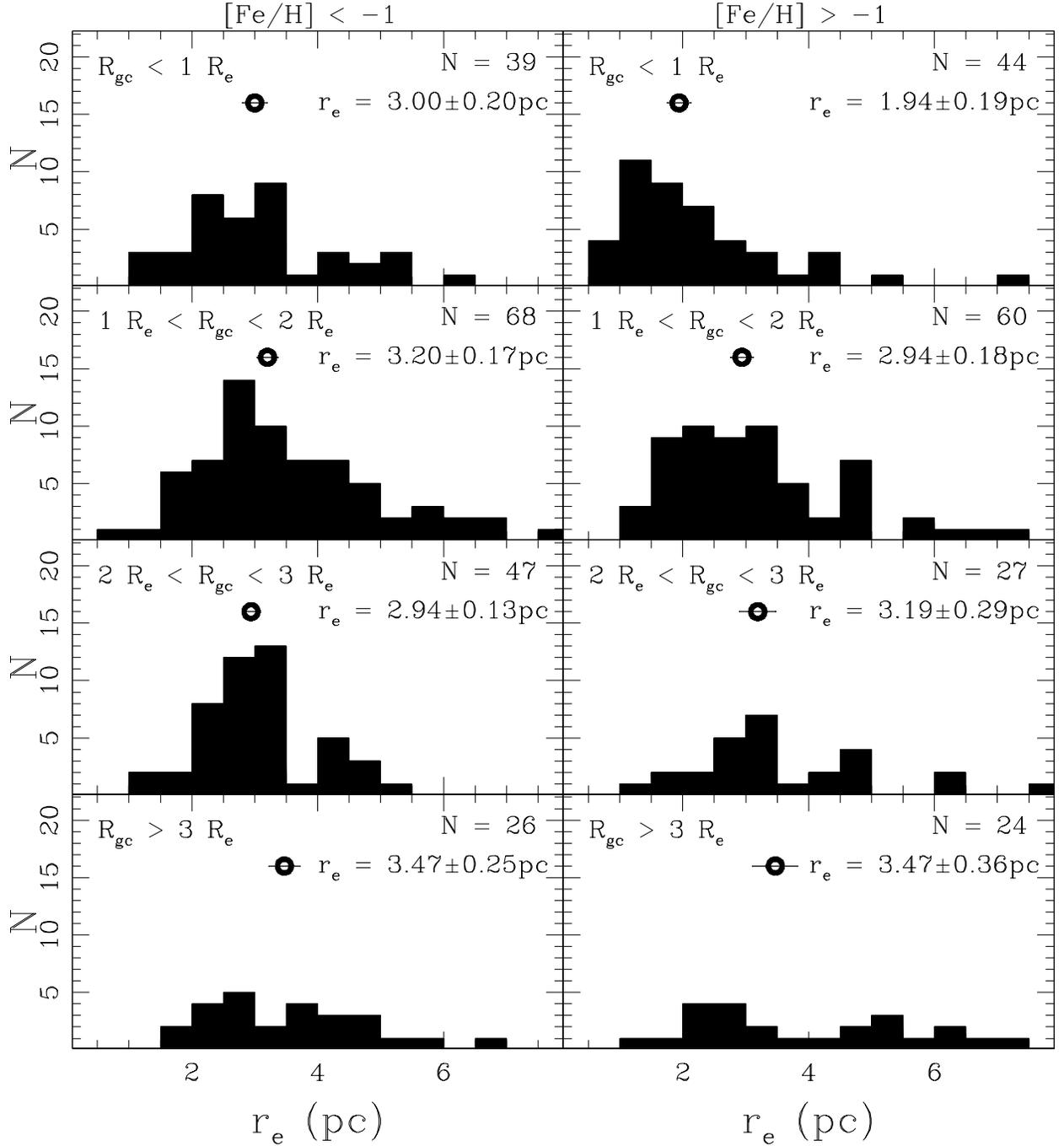}
\caption{The distributions in
  r$_e$ of the GCs in the same galactic effective radial
bins as in Fig.~\ref{fig:re_feh}, for metal-poor ([Fe/H]$ < -1$) on the left and
metal-rich ([Fe/H]$ > -1$) clusters on the right.  Associated number of
GCs and the mean r$_e$ (open circle) are labelled in
each bin.  Outside 3 R$_{eff}$ of the galaxy, the data on the GC
population suffers from incompleteness and spatial bias. The open circle
in each distribution gives the median value for the $r_e$ and its
formal error for each subsample. Metal-poor clusters show no significant
change of $r_e$ with R$_{gc}$, while, on average,  metal-rich clusters 
follow a clear trend
for larger $r_e$ with R$_{gc}$.} 
\label{fig:N_re}
\end{figure}

Figures~\ref{fig:re_feh} and \ref{fig:N_re} show that, within one effective radius of the
galaxy, the red clusters are significantly smaller than the blue
ones by $\sim 30\%$. 
As galactocentric distance increases, however, this 
difference tends to disappear and beyond a distance of 12 kpc (corresponding to
$\sim 2.3$ R$_{eff}$), no difference remains. We have performed a Spearman
test to study the trend between metallicity and r$_{e}$. The Spearman 
non-parametric rank-correlation coefficient ranges from -1 to 1, with -1 
for a complete anti-correlation, 0 for no correlation and 1 for a complete
correlation. We found the coefficients to be -0.4 for R$_{gc} < 1$ R$_{eff}$, 
-0.1 for 1 R$_{eff} <$ R$_{gc} <$ 2 R$_{eff}$, 0.1 for 2$ R_{eff} <$ R$_{gc} < 3 $R$_{eff}$, and
0.1 for R$_{gc} > 3$ R$_{eff}$.  In addition, the Spearman test gives the probability that
the two datasets are uncorrelated as well. The values read 0.01, 0.19, 0.43
and 0.34, respectively, for the four radial bins presented above. Thus,
there is a small, but virtually confirmed anti-correlation within 1 R$_{eff}$.

This size trend is even more clear in Figure~\ref{fig:N_re}.  
Within 1 R$_{eff}$ of the galaxy, the metal-rich and metal-poor clusters have median effective radii of
1.94$\pm0.19$ and 2.98$\pm0.20$ pc, respectively. 
However, outside of 1 R$_{eff}$, the 
median r$_e$ is identical within uncertainties. 

\begin{figure}
\plotone{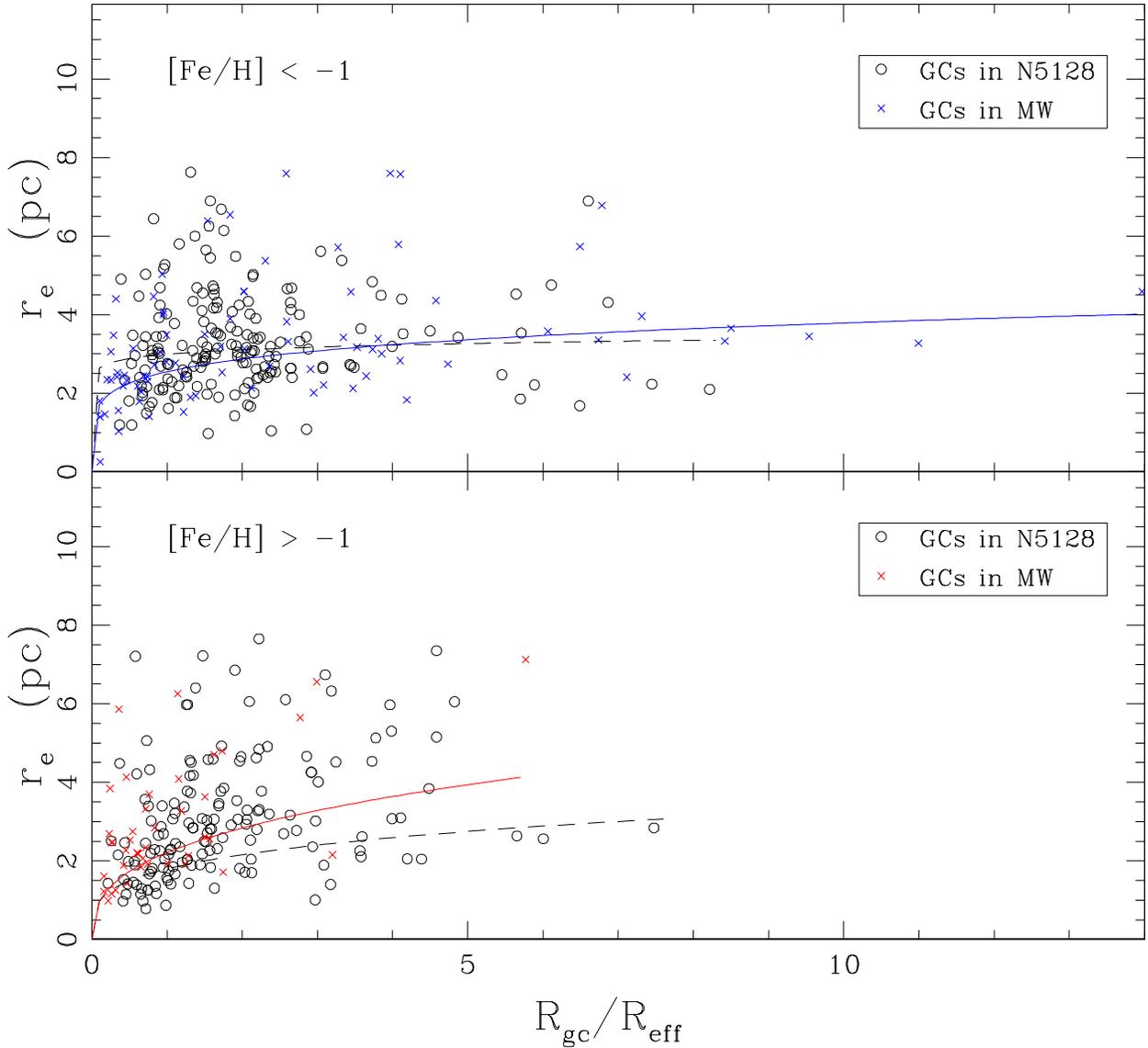}
\caption{The effective radius, r$_e$ in pc, as a function of projected
  galactocentric radius, R$_{gc}$ in kpc, for the GCs in NGC 5128 
  (open circles) and in the Milky Way galaxy (crosses) for both the 
  metal-poor GCs ([Fe/H]$ < -1$) on the top and metal-rich GCs 
  ([Fe/H]$ > -1$) on the bottom.  The Milky Way GC data is taken 
  from \cite{harris96} with the projected galactocentric radius defined as
  R$_{gc} = \sqrt{y^2 + z^2}$ and R$_{eff}=2.7$ kpc
  \citep{devaucouleurs78}.   
  Best fit curves of the form 
  r$_e =$c(R$_{gc}$/R$_{eff}$)$^\alpha$ yield (c$=3.22\pm0.12$, $\alpha=0.05\pm0.05$)
  and (c$=2.76\pm0.14$, $\alpha=0.26\pm0.06$) for the metal-poor 
  and metal-rich GCs in NGC 5128 (dashed curves)
  along with (c$=3.02\pm0.17$, $\alpha=0.17\pm0.04$) and 
  (c$=3.16\pm0.21$, $\alpha=0.36\pm0.07$) for the
  metal-poor and metal-rich GCs in the Milky Way (solid curves).}
\label{fig:re_Rgc}
\end{figure}

Figure~\ref{fig:re_Rgc} shows r$_{e}$ as a function of projected
R$_{gc}$ for the GCs in NGC 5128 and in the Milky Way.  
The metal-poor GCs in NGC 5128 do not follow the
r$_e$--R$_{gc}$ relationship that is evident in the metal-rich
GCs.  The Milky Way data, on the other hand, has been shown to host
the 
r$_e$--R$_{gc}$ relationship for both metallicity populations,
first noted by \cite{vandenbergh91}, using a 3-dimensional R$_{gc}$.  
However, in projection, the metal-poor GCs in the Milky Way
do not appear vastly different from those in NGC 5128.   

\section{Discussion}
\label{sec:discussion}
The existence of a systematic difference in the effective radii of 
blue and red clusters has been extensively studied in other galaxies,
with blue clusters typically 
found to be $17-30\%$ larger than red ones (see
Section~\ref{sec:intro}). However, in NGC 5128, \cite{harris02} did 
not find any correlation between color and size for a sample of 27 GCs using 
HST/WFPC2. In a subsequent study, \cite{gomez06} found the red GCs
to have {\em larger\/} median sizes compared to the blue clusters for a
sample of 38 objects, with Magellan/MagIC.
Both studies were based on small field images, involved small
sample sizes, and were centered at large R$_{gc}$. (At the
{\it smallest} R$_{gc}$, this was more than 2 times farther
than the R$_{eff}$ of the galaxy light.)
According to \cite{larsen03}, the average sizes
of red and blue clusters should be similar at about 1 R$_{eff}$ and
beyond,  if projection
effects are to account for the size difference.
The results from \cite{harris02} and \cite{gomez06} are 
consistent with this scenario, bearing in mind the 
low number statistics of their studies.

The only two studies thus far with a large enough sample
of GCs that also extends beyond 1 R$_{eff}$, are this work and
\cite{spitler06}.  Both show that the red clusters are smaller than the blue
within 1 R$_{eff}$, {\it and that they are identical in size beyond this
distance}.  In the ACS Virgo Cluster Survey, \cite{jordan05} have studied the
sizes of GCs in 67 early-type galaxies.
Their analysis reaches about 3 times the effective
radius of the massive ellipticals studied and these GCSs dominate their sample. 
However, their results, favoring red clusters being consistently smaller than
blue clusters are dominated by the inner GCs in these galaxies.

Our sample, consists of only confirmed GCs which have been analyzed
homogeneously, span a projected galactocentric
distance of up to 50 kpc, i.e., 8 galaxy R$_{eff}$.
Thus, we are able to draw conclusions about the origin of the size difference
with a sample that is both uncontaminated and much more spatially
extended than in previous studies.

As is evident from Figs.~\ref{fig:re_feh}, \ref{fig:N_re} and \ref{fig:re_Rgc} 
metal-poor clusters do not show an r$_e$--R$_{gc}$
relationship. \cite{jordan05} analyse this trend for metal-poor clusters
in their samples and conclude that they are too shallow compared to the
Galaxy for the projection effects to account for the size difference. Our
results for NGC 5128 agree with this, but at the same time make it clear that
the metal-poor subpopulation does not represent the global r$_e$--R$_{gc}$ trend.
In fact, only {\em metal-rich} clusters show it.
Therefore, projection effects can account for the observed size
differences without the need of intrinsic formation and destruction mechanisms
between red and blue clusters.

\section{Conclusions}
\label{sec:conclusions}

Using a contaminant-free sample of 364 GCs in NGC 5128,
confirmed with radial velocity measurements or by resolved HST images, 
we have measured effective radii using ISHAPE.  Our results indicate
that the blue or metal-poor clusters do not show any significant
r$_{e}$--R$_{gc}$ relation.  However, the red or metal-rich GCs 
do show a steep relation in which red clusters within 1
R$_{eff}$ of the galaxy's light are $30\%$ smaller than the blue
clusters.  Beyond this distance there is no indication for a size
difference between the two metallicity populations.  This finding in
NGC 5128, not previously seen in any other early-type galaxy, supports
the more tentative findings of the Sombrero galaxy's GCS 
\citep{spitler06}.  Both
studies support the idea that the size differences are most likely caused by
projection effects \citep{larsen03}
and not by intrinsic physical differences between the two subgroups.

Acknowledgements: M.G. and K.A.W. thank Dean McLaughlin for use of HST
structural parameters in advance of publication.  M.G. thanks the Dept. of 
Physics and Astronomy at McMaster University and especially Bill and 
Gretchen Harris for their hospitality.  K.A.W. thanks NSERC and
Bill Harris for financial support, and also the Depto. de
F{\'i}sica at the Universidad de Concepci{\'o}n, especially Doug
Geisler, for their hospitality. We thank the anonymous referee for
her/his valuable suggestions and comments.



\clearpage

\clearpage

\clearpage

\begin{thebibliography}{}


\bibitem[Aarseth \& Heggie(1998)]{aarseth98} Aarseth, S.~J. \& Heggie,
    D.~C. 1998, \mnras, 297, 794

\bibitem[Barmby et al.(2002)]{barmby02} Barmby, P., Holland, S., \&
  Huchra, J.~P. 2002, \aj, 123, 1937

\bibitem[Barmby et al.(2007)]{barmby07} Barmby, P., McLaughlin, D.~E.,
  Harris, W.~E., Harris, G.~L.~H., \& Forbes, D.~A. 2007, \aj, 133, 2764

\bibitem[Baggett et al.(1998)]{baggett98} Baggett, W.~E., Baggett, S.~M., \& Anderson, K.~S. 1998, \aj, 116, 1626

\bibitem[Brodie \& Strader(2006)]{brodie06} Brodie, J.~P. \&
Strader, J. 2006, \araa, 44, 193

\bibitem[C{\^o}t{\'e} et al.(2001)]{cote01} C{\^o}t{\'e}, P.,
  McLaughlin, D.~E., Hanes, D.~A., Bridges, T.~J., Geisler, D.,
  Merritt, D., Hesser, J.~E., Harris, G.~L.~H., \&  Lee, M.~G. 2001,
  \apj, 559, 828

\bibitem[de Vaucouleurs \& Pence(1978)]{devaucouleurs78} de
  Vaucouleurs, G. \& Pence, W.~D. 1978, \aj, 83, 1163

\bibitem[Dirsch et al.(2003)]{dirsch03} Dirsch, B., Richtler, T.,
  Geisler, D., Forte, J.~C., Bassino, L.~P., \& Gieren, W.~P. 2003, \aj,
  125, 1908

\bibitem[Djorgovski(1995)]{djorgovski95} Djorgovski, S. 1995, \apj,
  438, 29

\bibitem[Drinkwater et al.(2000)]{drinkwater00} Drinkwater, M.~J., Jones, J.~B., Gregg, M.~D., \& Phillipps, S. 2000, PASA, 17, 227

\bibitem[G{\'o}mez et al.(2006)]{gomez06}G{\'o}mez, M., Geisler, D., Harris, W.~E., Richtler, T.,
    Harris, G.~L.~H., \& Woodley, K.~A. 2006, \aap, 447, 877

\bibitem[Harris et al.(2004)]{harris04} Harris, G.~L.~H., Harris,
  W.~E., \& Geisler, D. 2004, \aj, 128, 723 

\bibitem[Harris(1996)]{harris96} Harris, W.~E. 1996, \aj, 112, 1487

\bibitem[Harris \& Harris(2002)]{harrisharris02} Harris, W.~E., \& Harris, G.~L.~H. 2002, \aj, 123, 3108

\bibitem[Harris et al.(2002)]{harris02} Harris, W.~E., Harris,
  G.~L.~H., Holland, S.~T., \& McLaughlin, D.~E. 2002, \aj, 124, 1435

\bibitem[Harris et al.(2006)]{harris06} Harris, W.~E., Harris,
  G.~L.~H., Barmby, P., McLaughlin, D.~E., \& Forbes, D.~A. 2006, \aj,
  132, 2187


\bibitem[Jord{\'a}n (2004)]{jordan04} Jord{\'a}n, A. 2004, \apj, 613L, 117

\bibitem[Jord{\'a}n et al.(2005)]{jordan05} Jord{\'a}n, A.,
  C{\^o}t{\'e}, P., Blakeslee, J.~P., Ferrarese, L., McLaughlin, D.~E.,
  Mei, S., Peng, E.~W., Tonry, J.~L., Merritt, D., Milosavljevi{\'c},
  M., Sarazin, C.~L., Sivakoff, G.~R., \& West, M.~J. 2005, \apj, 634, 1002

\bibitem[King (1962)]{king62} King, I.~R. 1962, \aj, 67, 471

\bibitem[Kundu \& Whitmore(1998)]{kundu98} Kundu, A. \& Whitmore,
  B.~C. 1998, \aj, 116, 2841

\bibitem[Kundu et al.(1999)]{kundu99} Kundu, A., Whitmore, B.~C.,
  Sparks, W.~B., Macchetto, F.~D., Zepf, S.~E., \& Ashman, K.~M. 1999,
  \apj, 513, 733

\bibitem[Kundu \& Whitmore(2001)]{kundu01} Kundu, A. \& Whitmore,
  B.~C. 2001, \aj, 121, 2950

\bibitem[Larsen (1999)]{larsen99} Larsen, S.~S. 1999, \aaps, 139, 393

\bibitem[Larsen (2001)]{larsen_s01} Larsen, S.~S. 2001, \aj, 122, 1782

\bibitem[Larsen et al.(2001a)]{larsen01} Larsen, S.~S., Brodie, J.~P., Huchra, J.~P., Forbes, D.~A., \& Grillmair, C.~J. 2001a, \aj, 121, 2974

\bibitem[Larsen et al.(2001b)]{larsen_fb01} Larsen, S.~S., Forbes, D.~A., \&
  Brodie, J.~P. 2001b, \mnras, 327, 1116

\bibitem[Larsen \& Brodie(2003)]{larsen03} Larsen, S.~S. \& Brodie,
  J.~P. 2003, \apj, 593, 340

\bibitem[McLaughlin(2000)]{mclaughlin00} McLaughlin, D.~E. 2000, \apj,
  539, 618

\bibitem[McLaughlin \& van der Marel(2005)]{mclaughlin05} McLaughlin,
D.~E. \& van der Marel, R.~P. 2005, \apjs, 161, 304

\bibitem[McLaughlin et al.(2007)]{mclaughlin07} McLaughlin, D.~E.,
  Barmby, P., Harris, W.~E., Harris, G.~L.~H., \& Forbes, D.~A. 2007, \mnras, submitted

\bibitem[Puzia et al.(1999)]{puzia99} Puzia, T.~H., Kissler-Patig, M.,
  Brodie, J.~P., \& Huchra, J.~P. 1999, \aj, 118, 2734 

\bibitem[Schlegel et al.(1998)]{schlegel98} Schlegel, D.~J., Finkbeiner, D.~P., \& Davis, M. 1998, \apj, 500, 525

\bibitem[Spitler et al.(2006)]{spitler06} Spitler, L.~R., Larsen,
  S.~S., Strader, J., Brodie, J.~P., Forbes, D.~A., \& Beasley, M.~A,
  2006, \aj, 132, 1593

\bibitem[Spitzer \& Thuan(1972)]{spitzer72} Spitzer, L.~Jr., Thuan,
  T.~X. 1972, \apj, 175, 31

\bibitem[van den Bergh et al.(1991)]{vandenbergh91} van den Bergh, S.,
  Morbey, C., \& Pazder, J. 1991, \apj, 375, 594

\bibitem[Woodley et al.(2005)]{woodley05} Woodley, K.~A., Harris, W.~E.,
  \& Harris, G.~L.~H, 2005, \aj, 129, 2654

\bibitem[Woodley et al.(2007)]{woodley07} Woodley, K.~A., Harris,
  W.~E., Beasley, M.~A., Peng, E.~W., Bridges, T.~J., Forbes, D.~A.,
  \& Harris, G.~L.~H 2007, \aj, 134, 494
   
\end{thebibliography}
\end{document}